\documentclass[preprint]{elsarticle}
\pdfoutput=1
\usepackage{amsmath}
\usepackage{amssymb}
\usepackage[final]{epsfig}
\usepackage{color}
\usepackage{graphicx}
\usepackage{lineno}
\newcommand{\beq}{\begin{equation}}
\newcommand{\eeq}{\end{equation}}
\newcommand{\bdi}{\begin{displaymath}}
\newcommand{\edi}{\end{displaymath}}
\newcommand{\no}{\nonumber}
\newcommand{\bea}{\begin{eqnarray}}
\newcommand{\eea}{\end{eqnarray}}
\newcommand{\vep}{\varepsilon}

\newcommand{\vx}{\vec{x}}

\newcommand{\di}{\vec{\nabla}}

\newcommand{\de}{\partial}

\begin{document}
\begin{frontmatter}

\title{An application of extensions of the Ramo-Shockley theorem to signals in silicon sensors}

\author[CERN]{W. Riegler}

\address[CERN]{CERN, Geneva, Switzerland}

\begin{abstract}

We discuss an extension of the Ramo-Shockley theorem that allows the calculation of signals in detectors that contain non-linear materials of arbitrary permittivity and finite conductivity (volume resistivity) as well as a static space-charge. The readout-electrodes can be connected by an arbitrary impedance network. This formulation is useful for the treatment of semiconductor sensors where the finite volume resistivity in the sensitive detector volume cannot be neglected. The signals are calculated by means of time dependent weighting fields and weighting vectors. These are calculated by adding voltage or current signals to the electrodes in question, which has a very practical application when using semiconductor device simulation programs. An analytic example for an un-depleted silicon sensor is given.


\end{abstract}

\end{frontmatter}

\section{Introduction}

The currents induced on grounded electrodes by  moving charges can be calculated with static weighting fields using the Ramo-Shockley  
theorem \cite{ramo}\cite{shockley}. The extension of the theorem for the presence of a static space-charge in silicon sensors is treated in \cite{gatti}. In case the electrodes are not grounded but connected with linear impedance elements, the voltages and currents 
can be calculated by time dependent weighting fields as shown in \cite{radeka} or by application of an equivalent circuit diagram as shown in  \cite{blum}. The presence of dielectric and nonlinear media in the detector is treated in \cite{hamel1}\cite{hamel2}. The case where the volume between the electrodes contains conductive material is treated in \cite{werner1}\cite{werner2}. In this report we write the theorems presented in \cite{werner2} in a form that is very useful when calculating signals in a partially depleted silicon sensor with TCAD device simulation programs, as outlined in the following.
\\ \\
The theorems in \cite{werner2} were first applied to Resistive Plate Chambers (RPCs) \cite{werner1}, where the effect of the finite resistivity of the plates on the signals was investigated. The volume resistivity of materials used for RPCs ranges from $10^{10}-10^{12}\,\Omega$cm and it is independent of the applied voltage. In silicon sensors the volume resistivity does however depend on the applied voltage, which is why we refer to it as a 'non-linear' material. Using a TCAD device simulation program we can define a sensor geometry with a given doping profile and apply the bias voltages to find the static electric field and the density of electrons $n_e(\vx)$ and holes $n_h(\vx)$ in the sensor volume. The conductivity $\sigma(\vx)$, which is the inverse of the volume resistivity, is then 
\beq
    \sigma (\vx) = q [\mu_e n_e(\vx) + \mu_h n_h(\vx)]
\eeq
where $q$ is the elementary charge, $\mu_e$ is the electron mobility and $\mu_h$ is the hole mobility. In order to be consistent with \cite{werner2} we will use the conductivity $\sigma$ instead of the volume resistivity in the following. In case the sensor is fully depleted we have $n_e=n_h=0$ and therefore $\sigma=0$ and the standard Ramo-Shockley theorem using static weighting fields can be applied. In case a silicon sensor is only partially depleted, the finite conductivity $\sigma(\vx)$ of the detector volume will influence the induced signal and the time dependent weighting fields and weighting vectors have to be used. To calculate them according to \cite{werner2} one has to ground all electrodes and apply a delta current or delta voltage to the electrode in question. Performing this calculation with a TCAD device simulation program will however not yield the correct result, since for this electrostatic arrangement  the detector is completely unbiased and does not have the correct distribution of conductivity. There are two ways to perform the calculation:
\begin{itemize}

    \item    One takes the simulated distribution of conductivity $\sigma(\vx)$ into a separate calculation and applies the theorems as outlined in \cite{werner2}.

    \item    One adds a small voltage or current pulse to the electrode in question for the correctly biased sensor and takes the difference of the resulting time dependent field and the static field.

\end{itemize}
Both of these recipes will yield the same result as shown for the case of static weighting fields in \cite{hamel2} and as will be outlined for the time dependent weighting fields in the next section.
\\ \\
 The method of weighting fields is only applicable if the electric field due to the charge deposited in the silicon sensor has negligible impact on the electron and hole density in the sensor. In that case the weighting field can be imagined as the 'linearization' of the problem around the bias points. For very large charge deposits where this condition is not satisfied, the problem becomes nonlinear and the signal can only be calculated by iterating the full field calculation in the sensor, which is rather time consuming. 
\\
The big advantage of the weighting field method is the computational efficiency for Monte Carlo simulations. Once the time dependent weighting field is calculated with TCAD, the charge deposit, transport of charges and calculation of the induced signal can then be performed with programs like Garfield \cite{garfield1}\cite{garfield2} by convoluting the velocity vectors of the moving charges with the time dependent weighting fields. Another advantage is that such a calculation takes the nature of the electrons and holes as being point charges into account. In some TCAD simulations the charge deposit has to be formulated as a continuous functional distribution and therefore also single electrons and holes will 'diffuse', which is unphysical.
\\ \\
In the following we first discuss the weighting field theorems as outlined above and then we discuss a simple example of an un-depleted silicon sensor that can be used as a benchmark example for a TCAD simulation.

\section{Theorems}

\begin{figure}[h]
\begin{center}
     \epsfig{file=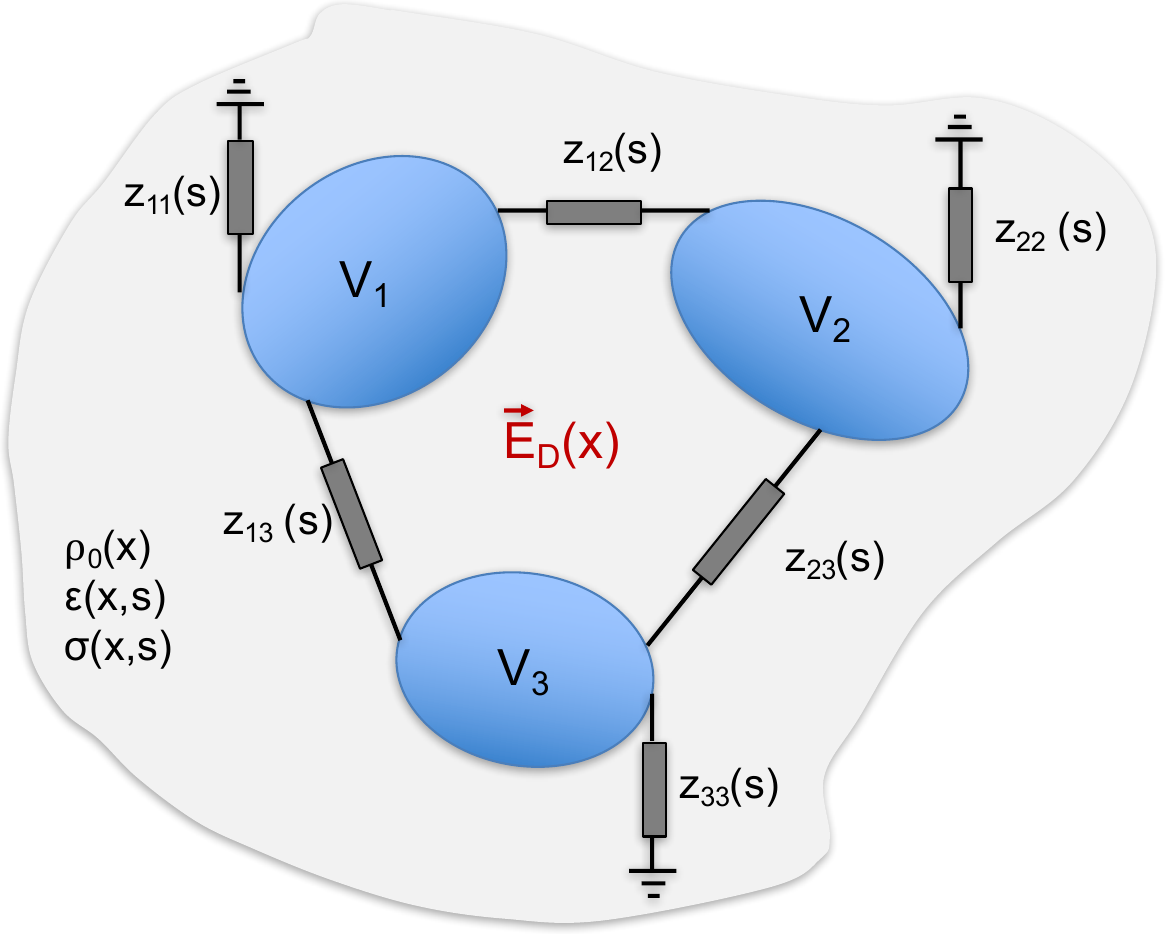, height=5cm}
\caption{The electric field $\vec E_D(\vx)$ in the detector volume that is responsible for the movement of the charge. The Laplace parameter $s=i\omega$ represents the frequency dependence of the material properties and the impedance elements. }
\label{driftfield}
\end{center}
\end{figure}

\noindent
In Fig. \ref{driftfield}  we see a system of electrodes at static voltages $V_n$ embedded in a medium with permittivity $\vep(\vx)$, conductivity $\sigma(\vx)$ and space-charge $\rho_0(\vx)$. In case $\sigma=0$ the medium represents an insulator and the static potential $\phi_D(\vx)$ is given by
\beq
    \phi_D(\vx) = \sum_{n=1}^N \frac{V_n}{V_0}\psi_n (\vx) + \phi_0(\vx)
\eeq
where $\psi_n(\vx)$ are the static weighting potentials of the electrodes and $\phi_0(\vx)$ is the potential due to $\rho_0(\vx)$ in case all electrodes are grounded. This solution satisfied the boundary conditions by construction, since the weighting potential $\psi_n(\vx)$ is equal to $V_0$ on the surface of electrode $n$ and zero on all other electrodes and therefore $\phi_D(\vx)\vert _{\vx = \vec{A}_n}=V_n$ is guaranteed. Since a solution for a given boundary condition is unique, this is the correct solution. Now, adding the voltage $V_0$ to $V_1$ results in the potential
\beq
    \phi_{D1}(\vx) = \psi_1(\vx)+\sum_{n=1}^N \frac{V_n}{V_0}\psi_n (\vx) + \phi_0(\vx) 
\eeq
so we have
\beq
     \psi_1(\vx) =    \phi_{D1}(\vx) - \phi_D(\vx) 
\eeq
\\ \\
The weighting field of an electrode can therefore also be calculated by leaving all electrodes at the bias voltages and adding the voltage $V_0$ to the electrode in question and then taking the difference of the fields \cite{hamel2}. Since this argument holds for solutions of the Poisson equation it also applies to the time dependent fields when using the quasi-static approximation.
From these considerations the theorems presented in \cite{werner2}, specifically Eqs. (3)(7)(14)(18), will be formulated in a more general way in the following. We will make frequent use of the Heaviside step function $\Theta (t)$ which is zero for $t<0$ and unity for $t \ge 0$.

\newpage

\subsection{{\bf Induced voltage}}

\begin{figure}[h]
\begin{center}
a)
     \epsfig{file=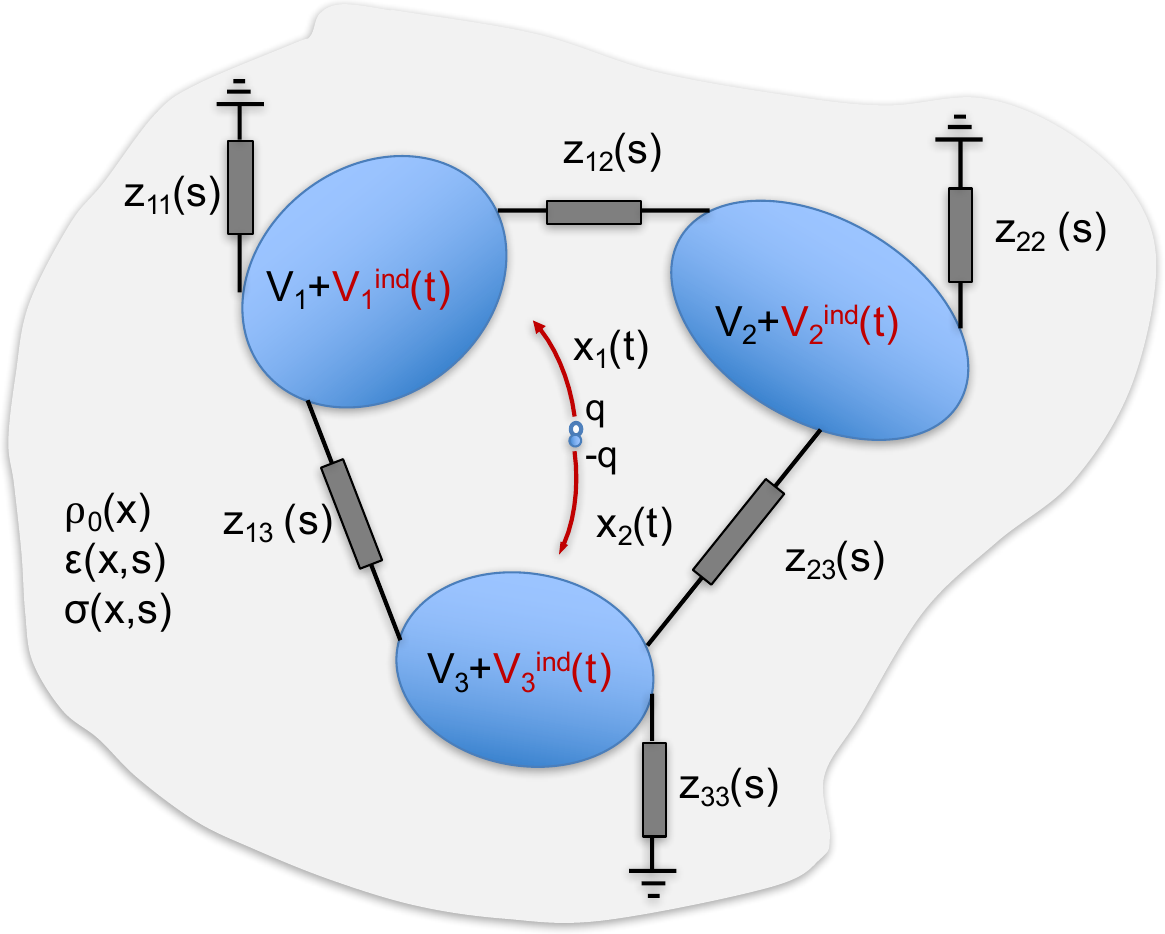, height=5cm}
b)
\epsfig{file=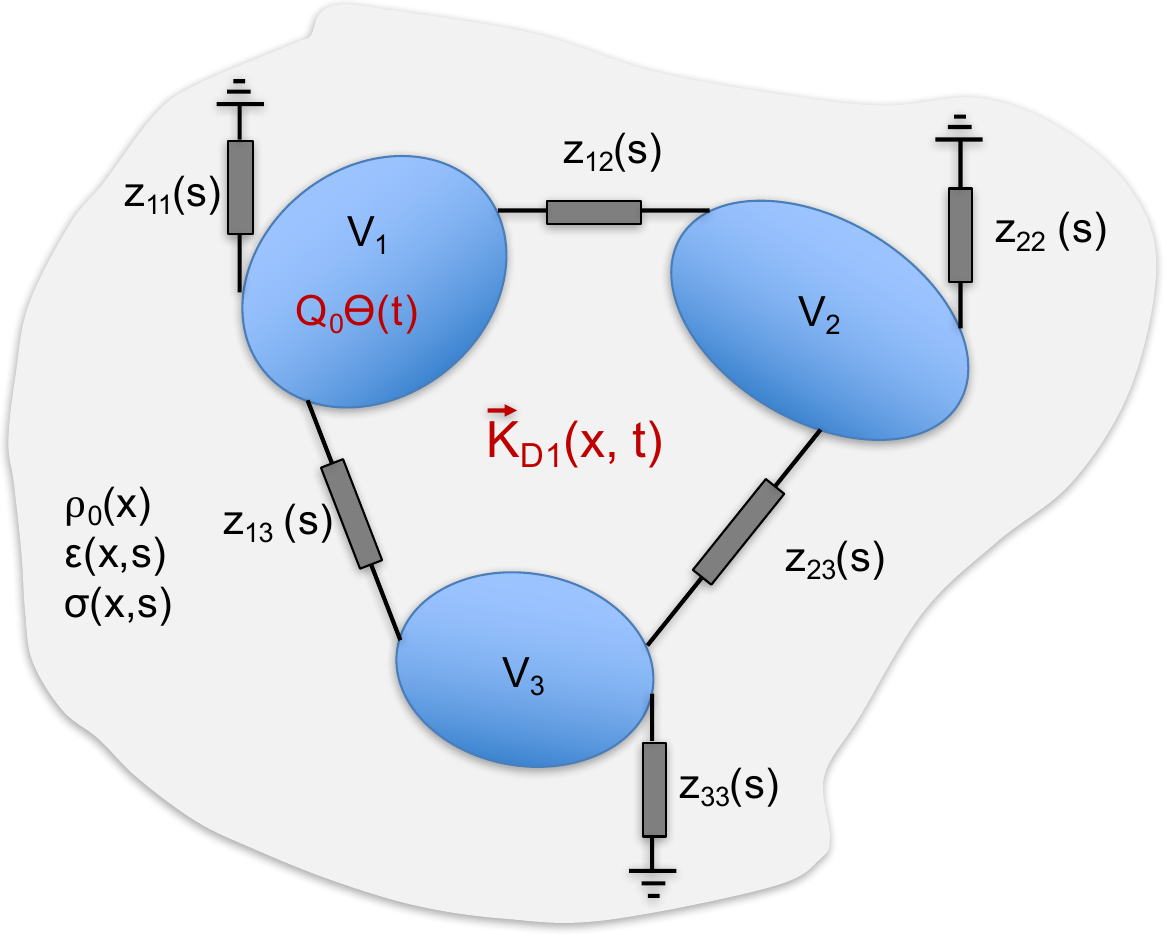, height=5cm}
\caption{ a) Two point charges $q, -q$ placed in the detector at $t=0$. Their movement induces voltages $V_n(t)$ on the electrodes.  b)  The electric field $\vec{K}_{D1}(\vx, t)$ due to placing a charge $Q(t)=Q_0\Theta(t)$ on electrode 1. This is equivalent to placing a current $I(t)=Q_0 \delta(t)$ on the electrode.}
\label{weightfield1}
\end{center}
\end{figure}
\noindent
\\
{\bf Theorem 1}
\\
{\it 
A pair of charges $q, -q$ is created in the detector at $t{=}0$ and these charges are moving in the electric field $\vec E_D(\vx)$ along trajectories $\vx_1(t)$ and $\vx_2(t)$. The voltage $V^{ind}_n(t)$ induced on electrode $n$  (Fig.\,\ref{weightfield1}a) can be calculated by }
\begin{eqnarray} \label{induced_voltage}
    V^{ind}_n(t) &=& 
     -\frac{q}{Q_0}\int_0^t \vec{K}_n\left[\vec{x}_1(t'),t-t'\right] 
    \vec{\dot{x}}_1(t') dt'  \\ 
     && + \frac{q}{Q_0}\int_0^t 
    \vec{K}_n\left[\vec{x}_2(t'),t-t'\right] \vec{\dot{x}}_2(t') dt'  \nonumber
\end{eqnarray}
{\it where  the weighting field $\vec K_n(\vx, t)$ is defined the following way (Fig.\,\ref{weightfield1}b): the charges $q,-q$ are removed, an 'infinitesimal'  charge $Q(t)=Q_0 \Theta(t)$ is added to electrode $n$, which results in a field $\vec K_{Dn}(\vx ,t)$, from which the weighting field $\vec K_n(\vx, t)$ is derived as }
\beq
   \vec{K}_n(\vx, t) = \vec{K}_{Dn}(\vx,t) - \vec{E}_{D}(\vx)
\eeq
 Since $I(t) = d (Q_0 \Theta(t))/dt = Q_0 \delta(t)$, the weighting field can be understood as being the result of a delta current pulse on the electrode in question. (In \cite{werner2} this was written as $I_0\delta(t)$ to indicate the delta current, which is however misleading since $\delta (t)$ has units of s$^{-1}$ and therefore only $Q_0\delta(t)$ has the correct units of Ampere.)

\newpage

\subsection{{\bf Induced charge on grounded electrode}}

\begin{figure}[h]
\begin{center}
a)
     \epsfig{file=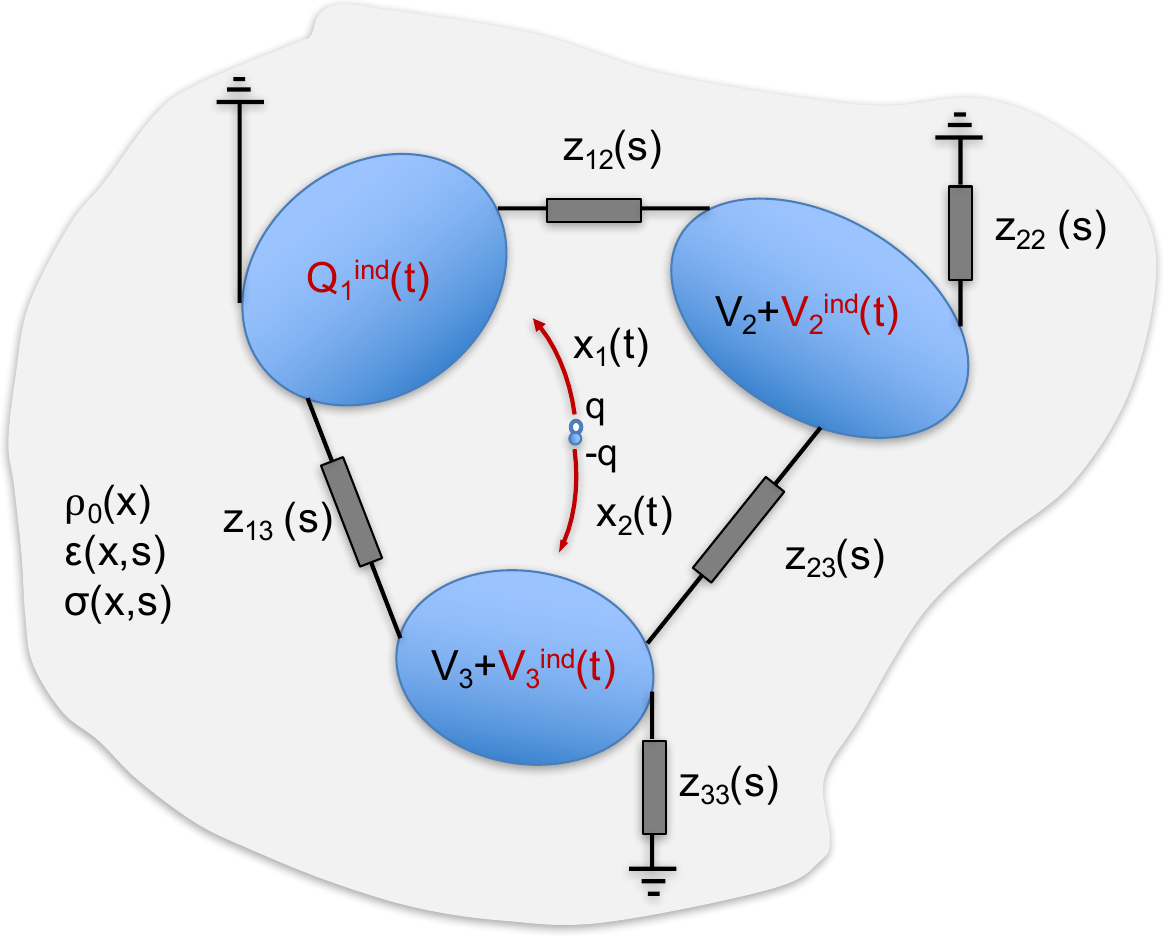, height=5cm}
b)
\epsfig{file=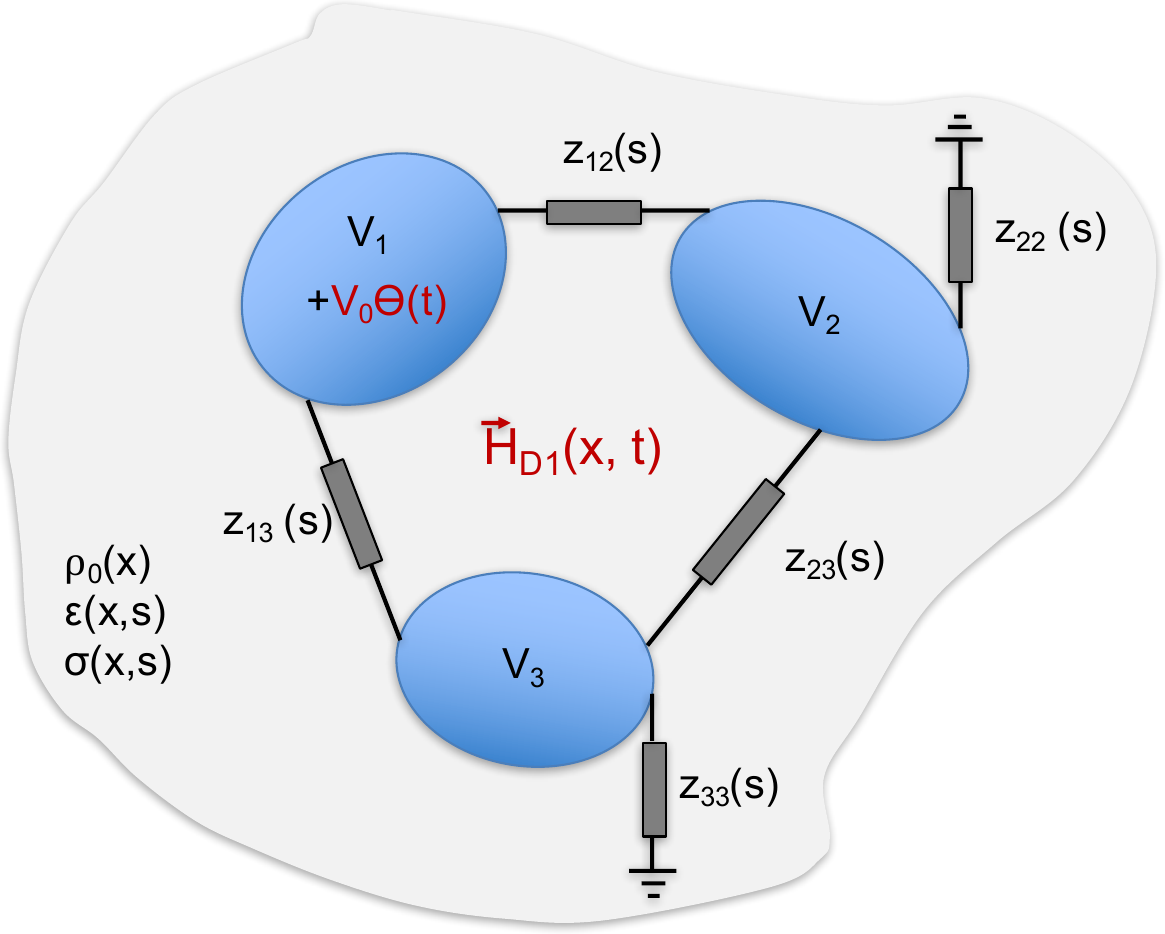, height=5cm}
\caption{ a) Two point charges $q, -q$ placed in the detector at $t=0$. Their movement induces a charge $Q_n(t)$ on a grounded electrode.  b)  The electric field $\vec{H}_{D1}(\vx, t)$ due to adding a voltage $V(t)=V_0\Theta(t)$ on electrode 1. }
\label{weightfield2}
\end{center}
\end{figure}
\noindent
In case an electrode is grounded or kept at a fixed potential, as shown in Fig.\,\ref{weightfield2}a), the voltage on the electrode stays unchanged and there is no 'induced voltage'. Still, the movement of the charges $q, -q$ induces a charge on this electrode, which we calculate in the following. Let us assume that electrode 1 is connected to ground through a very large capacitor $C$ such that the impedance $z_{11}=1/sC$ is negligible with respect to all other impedances in the circuit. The charge on this capacitor, specifically the charge on the capacitor plate on the 'ground side', is then $Q^{ind}_1(t) = -C V^{ind}(t)$. The charge $Q(t)=Q_0\Theta (t)$ that defines the weighting field will on the other hand result in a voltage $V(t)=Q_0/C \Theta (t) = V_0 \Theta (t)$ on the electrode. Identifying $Q_0/C$ with $V_0$ and taking the limit of $C \rightarrow \infty$ we then have the following theorem:
\\
\\
{\bf Theorem 2}
\\
{\it 
A pair of charges $q, -q$ is created in the detector at $t{=}0$ and these charges are moving in the electric field $\vec E_D(\vx)$ along trajectories $\vx_1(t)$ and $\vx_2(t)$. The charge $Q^{ind}_n(t)$ induced on the grounded electrode $n$  (Fig.\,\ref{weightfield2}a) can be calculated by }
\begin{eqnarray} \label{induced_charge}
    Q^{ind}_n(t) &=& 
     \frac{q}{V_0}\int_0^t \vec{H}_n\left[\vec{x}_1(t'),t-t'\right] 
    \vec{\dot{x}}_1(t') dt'  \\ 
     && - \frac{q}{V_0}\int_0^t 
    \vec{H}_n\left[\vec{x}_2(t'),t-t'\right] \vec{\dot{x}}_2(t') dt'  \nonumber
\end{eqnarray}
{\it where  the weighting field $\vec H_n(\vx, t)$ is defined the following way (Fig.\,\ref{weightfield2}b): the charges $q,-q$ are removed, an 'infinitesimal'  voltage $V(t)=V_0 \Theta(t)$ is added to electrode $n$, which results in a field $\vec H_{Dn}(\vx ,t)$, from which the weighting field $\vec H_n(\vx, t)$ is derived as }
\beq
   \vec{H}_n(\vx, t) = \vec{H}_{Dn}(\vx,t) - \vec{E}_{D}(\vx)
\eeq
Since we assume for all calculations the quasi-static approximation of Maxwell's equations, we do not consider propagation times of electric fields, so the application of the step voltage $V_0 \Theta(t)$ will result in an electric field $\Theta(t)\vec H_n(\vx, t) $ with a step at $t=0$ from where the field then evolves smoothly. In case all electrodes are grounded and the medium does not have any conductivity, the field $\vec H_n(\vx, t)$ does not have any time dependence beyond $t=0$. In that case we have
\bea
         Q^{ind}_n(t) &=& 
     \frac{q}{V_0}\int_0^t \vec{H}_n\left[\vec{x}_1(t'),0\right] 
    \vec{\dot{x}}_1(t') dt' 
      - \frac{q}{V_0}\int_0^t 
    \vec{H}_n\left[\vec{x}_2(t'),0\right] \vec{\dot{x}}_2(t') dt' \\
    & = & -\frac{q}{V_0} \psi_n[\vx_1(t)]  + \frac{q}{V_0} \psi_n[\vx_2(t)] \qquad \vec H_n(\vx, 0) = -\di \psi_n(\vx)  \nonumber 
\eea
The potential $\psi_n (\vx)$ is the weighting potential of electrode $n$ discussed earlier. 

\newpage

\subsection{{\bf Induced current on grounded electrode}}

\begin{figure}[h]
\begin{center}
 a)
 \epsfig{file=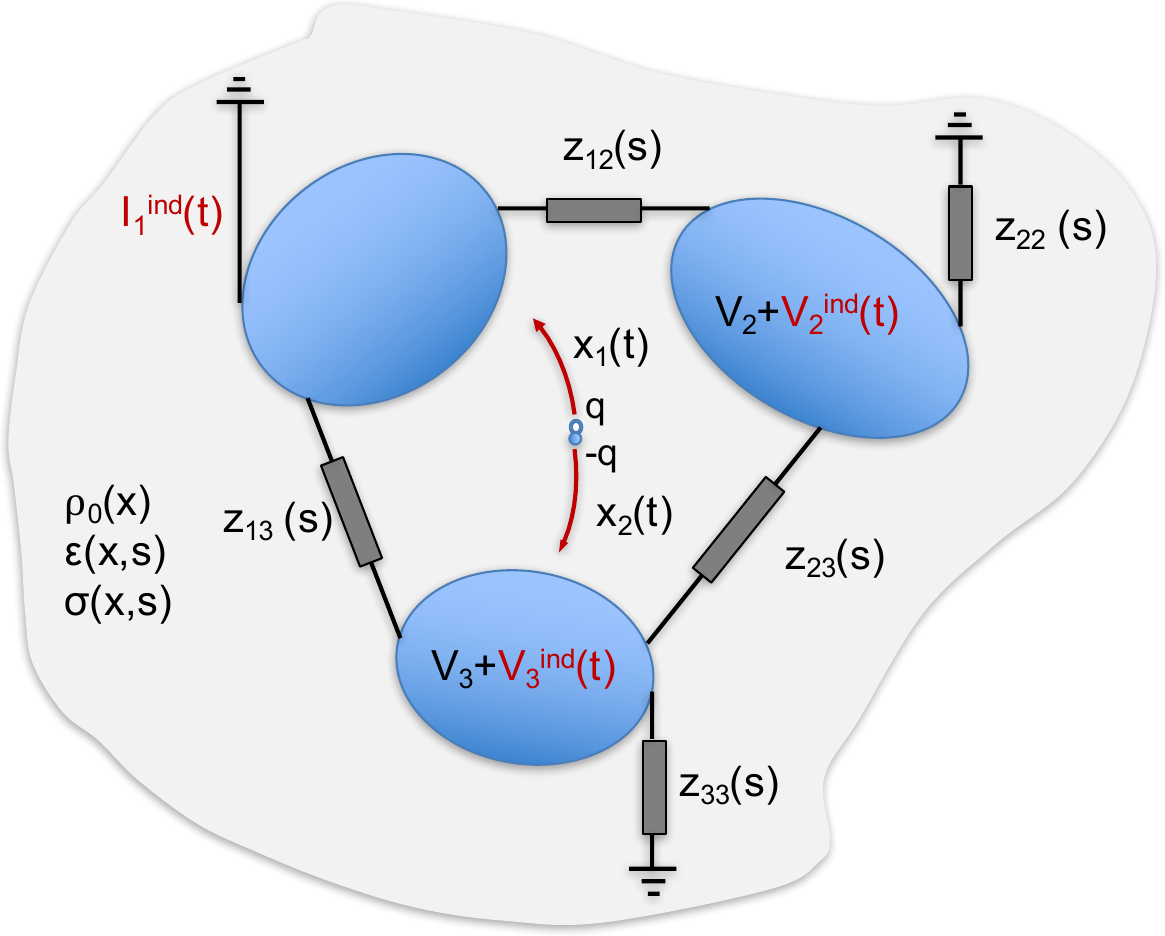, height=5cm}
 b)
\epsfig{file=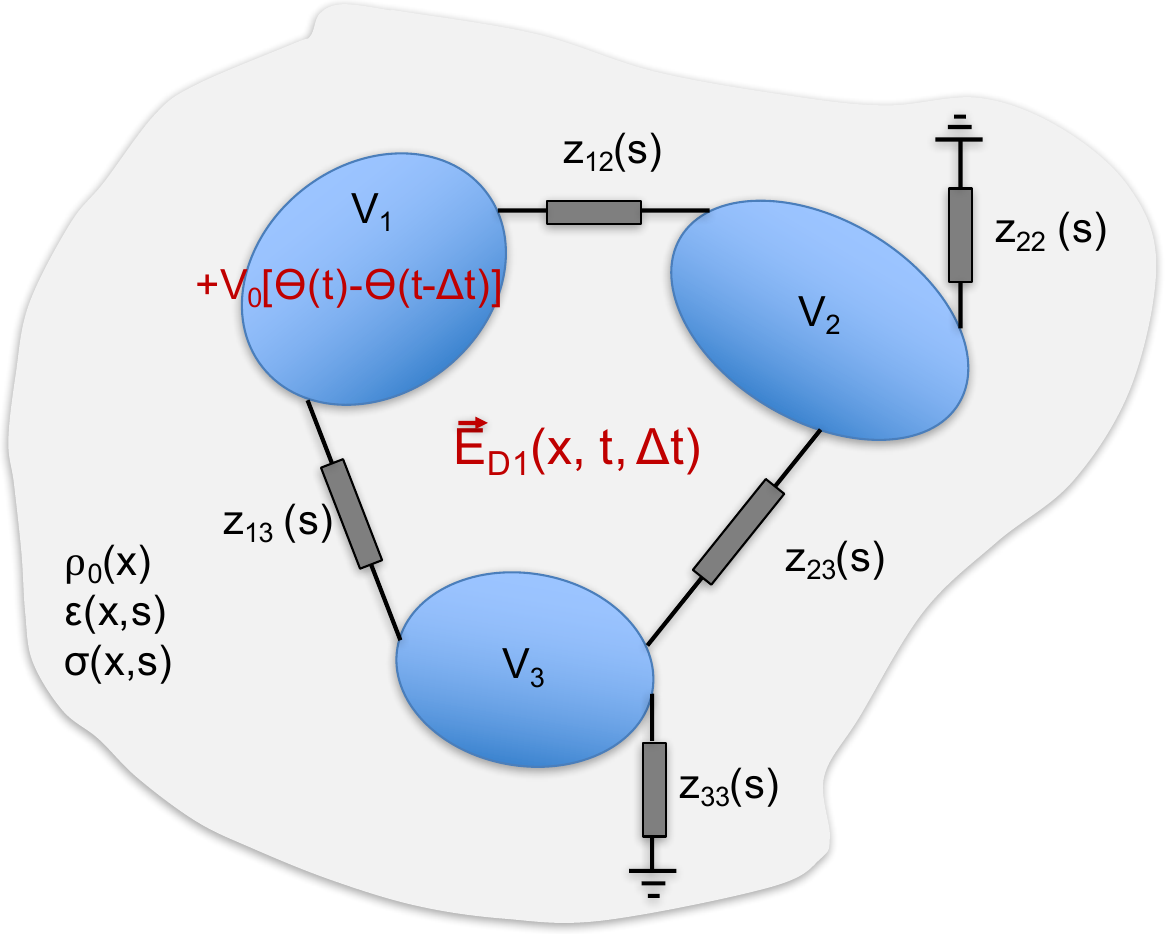, height=5cm}
\caption{a) Two point charges $q, -q$ placed in the detector at $t=0$. Their movement induces a current $I_n(t)$ on grounded electrodes b) The electric field $\vec{H}_{D1}(\vx, t, \Delta t)$ due to putting a 'square voltage pulse'  $V_0[\Theta(t)-\Theta(t-\Delta t)]$ on electrode 1.}
\label{weightfield3}
\end{center}
\end{figure}
\noindent
The current induced on a grounded electrode is simply defined as the derivative of the induced charge, which gives
\begin{eqnarray} \label{induced_current1}
    I^{ind}_n(t) &=& -\frac{d Q^{ind}_n(t)}{d t} \\
      & = & -  \frac{q}{V_0}
    \left[ 
    \vec{H}_n\left[\vec{x}_1(t),0\right]  \vec{\dot{x}}_1(t)  
    +\int_{0}^t \vec{H}^{(t)}_n\left[\vec{x}_1(t'),t-t'\right]  \vec{\dot{x}}_1(t') dt'       
     \right]  \nonumber \\ 
       &&
           + \frac{q}{V_0}
    \left[ 
    \vec{H}_n\left[\vec{x}_2(t),0\right]  \vec{\dot{x}}_2(t)  
    +\int_{0}^t \vec{H}^{(t)}_n\left[\vec{x}_2(t'),t-t'\right]  \vec{\dot{x}}_2(t') dt'       
     \right]  \nonumber
\end{eqnarray}
with $\vec H^{(t)}_n(\vx, t) = \de \vec H_n (\vx, t)/\de t$. Again, if all the electrodes are grounded and the medium has zero conductivity, the field $\vec H_n$ has no time dependence and the relation becomes
\beq
       I^{ind}_n(t) = -  \frac{q}{V_0} \vec{H}_n\left[\vec{x}_1(t),0\right]  \vec{\dot{x}}_1(t)  +  \frac{q}{V_0} \vec{H}_n\left[\vec{x}_2(t),0\right]  \vec{\dot{x}}_2(t) 
\eeq
which recuperates the Ramo-Shockley theorem with $\vec H_n(\vx) = - \di \psi_n(\vx)$. So we see that the reaction of the medium and the impedance network is all encoded in the time dependent term of the expression above. Using the relation
\bea
   &&  \frac{d}{d t} \int_0^t  \vec{H}_n\left[\vec{x}(t'),t-t'\right]  \vec{\dot{x}}(t') dt'  \\
      & = &  \int_0^\infty  \frac{d}{d t} \left( \Theta(t-t') \vec{H}_n\left[\vec{x}(t'),t-t'\right] \right) \vec{\dot{x}}(t') dt'   \nonumber \\
      & = &  \int_0^t \vec{W}_n[\vx(t'),t-t']\vec{\dot{x}}(t') dt'   \nonumber
\eea
we have defined the weighting vector $\vec W_n(\vx, t)$ as 
\beq
   \vec W_n(\vx, t)  =  \frac{\de}{\de t}\left[\Theta(t)\vec H_n(\vx, t)\right] \\
   =  \delta(t) \vec H_n(\vx, t) + \Theta(t) \frac{\de}{\de t} \vec H_n(\vx, t)
\eeq
so we see that this weighting vector has a 'prompt' and a 'delayed' component. Because it holds that
\bea
   \vec W_n(\vx, t) & = & \frac{\de}{\de t}\left[ \Theta(t)\vec H_n(\vx, t)  \right] \\
   & = & \lim_{\Delta t \rightarrow 0} \frac{1}{\Delta t}
   \left[
      \Theta(t)\vec H_n(\vx, t) -  \Theta(t-\Delta t)\vec H_n(\vx, t-\Delta t)
   \right]  \nonumber 
\eea
we see that the field  $\vec{W}_n(\vx, t)$ is the result of a square voltage pulse of amplitude $V_0$ and duration $\Delta t$, so we can state: \\
\\
{\bf Theorem 3}
\\
{\it 
A pair of charges $q, -q$ is created in the detector at $t{=}0$ and these charges are moving in the electric field $\vec E_D(\vx)$ along trajectories $\vx_1(t)$ and $\vx_2(t)$. The current $I^{ind}_n(t)$ induced on a grounded electrode (Fig.\,\ref{weightfield2}a) can be calculated by}
\begin{eqnarray} \label{induced_current2} 
    I^{ind}_n(t) &=& 
    -\frac{q}{V_0}\int_0^t \vec{W}_n\left[\vec{x}_1(t'),t-t'\right] 
    \vec{\dot{x}}_1(t') dt' \\
    &&  + \frac{q}{V_0}\int_0^t 
    \vec{W}_n\left[\vec{x}_2(t'),t-t'\right] \vec{\dot{x}}_2(t') dt' \nonumber 
\end{eqnarray}
{\it where  the weighting vector $\vec W_n(\vx, t)$ is defined the following way (Fig.\,\ref{weightfield3}b): the charges $q,-q$ are removed, an 'infinitesimal'  voltage pulse $V(t)=V_0 [\Theta(t)-\Theta (t- \Delta t)]$ is added to electrode $n$, which results in a field $\vec E_{Dn}(\vx ,t, \Delta t)$, from which the weighting vector $\vec W_n(\vx, t)$ is derived as }
\beq
      \vec{W}_n(\vx, t) =  \lim_{\Delta t \rightarrow 0} \frac{1}{\Delta t} [\vec{E}_{Dn}(\vx, t, \Delta t) - \vec{E}_{D}(\vx)]
\eeq
In contrast to $\vec K(\vx, t)$ and $\vec H(\vx, t)$, the vector $\vec W(\vx, t)$ does not represent an electric field but a vector with units V/(cm s). For the case where $E_D(\vx)=0$ the weighting vector $W_n(\vx, t)$ then becomes the response to a delta pulse $V_0\delta(t)$ on the electrode $n$.

\newpage

\subsection{\bf Application of the theorems}

When applying these theorems to silicon sensors, the pulses $V_0 \delta(t)$, $Q_0[ \Theta(t) - \Theta(t-\Delta t)]$ and $Q_0 \delta(t)$ have to be 'infinitesimal' or in practice chosen such that they do not alter the distribution of volume resistivity (conductivity) in the sensor.
The theorem for calculation of the induced voltage $V^{ind}_n(t)$ applies to the case where the impedance of the readout electronics and the biasing network cannot be neglected.
Since the detector 'sees' only the input impedance of an amplifier and not all internal details of the amplifier circuit, it is most practical to connect an element representing this input impedance to the electrode, then calculate the induced voltage and use this induced voltage as a source term for the detailed readout circuit simulation in a dedicated analog circuit simulation program. 
\\ 
The theorems for the induced charge $Q^{ind}_n(t)$ and the induced current $I^{ind}_n(t)$ apply in case the input impedance (resistance) of the readout electronics is negligible with respect to the electrode impedances and the impedance of the biasing network. 
\\
For practical application, when using e.g. TCAD simulations, one calculates the field $\vec E_{Dn}(\vx, t)$ by adding a 'short' voltage pulse $V(t)$ of arbitrary functional form and duration $T$ to the electrode in question, subtracts the static field $\vec E_D(\vx)$ and then divides the resulting electric field by $\int_0^T V(t) dt$ to find the equivalent of $\vec W_n(\vx, t)/V_0$. The length $T$ of the pulse has to be chosen to be smaller than the reaction time $\tau$ of the medium. For a homogenous medium with dielectric permittivity $\varepsilon_r\varepsilon_0$ and conductivity $\sigma$  this time corresponds to $\tau = \varepsilon_r \varepsilon_0/\sigma$. \\ 
To avoid numerical complications with the 'prompt' initial part of the weighting field it is useful to split the weighting field into the 'prompt' component and the 'delayed' component and use expression \ref{induced_current1} to calculate the signals. When adding the 'short' voltage pulse $V_0(t)$ to the electrode in a TCAD simulation, the electric field will be strictly proportional to the applied voltage for $t<T$ and then show the reaction of the medium for $t>T$. The expressions  of Eq. \ref{induced_current1} then have to be replaced by
\bea
     \frac{\vec H_n(\vx, 0)}{V_0} & \rightarrow & \frac{1}{V_0(t)} \left[ \vec E_{Dn}(\vx, t)  - \vec E_D(\vx)\right] \quad t<T \\ 
     \frac{ \vec H^{(t)}_n(\vx, t)}{V_0}  & \rightarrow &   \frac{1}{\int_0^TV_0(t)dt} \left[ \vec E_{Dn}(\vx, t)  - \vec E_D(\vx)\right] \quad t>T
\eea
The second expression has to then be extrapolated back to zero from $t>T$. If we use the simple example of a triangular pulse of duration $T$ with a peak of $V_0$ at $t=T/2$,  this corresponds to
\beq
     \frac{ \vec H_n(\vx, 0)}{V_0}   \rightarrow   \frac{1}{V_0} \left[ \vec E_{Dn}(\vx,T/2)  - \vec E_D(\vx)\right]   \qquad 
     \frac{\vec H^{(t)}_n(\vx, t)}{V_0} \rightarrow \   \frac{2}{T V_0} \left[ \vec E_{Dn}(\vx, t)  -\vec E_D(\vx)\right]
\eeq
In case one wants to avoid dealing with the 'prompt' and 'delayed' part of the weighting vector $\vec W_n (\vx, t)$ one can use the weighting field $\vec H_n(\vx, t)$ by applying in TCAD a voltage step function with a rise-time smaller than $T$, calculate the induced charge on the electrode and then perform the time derivative to find the induced current.


\newpage

\section{Examples}

\begin{figure}[h] 
\begin{center}
\epsfig{file=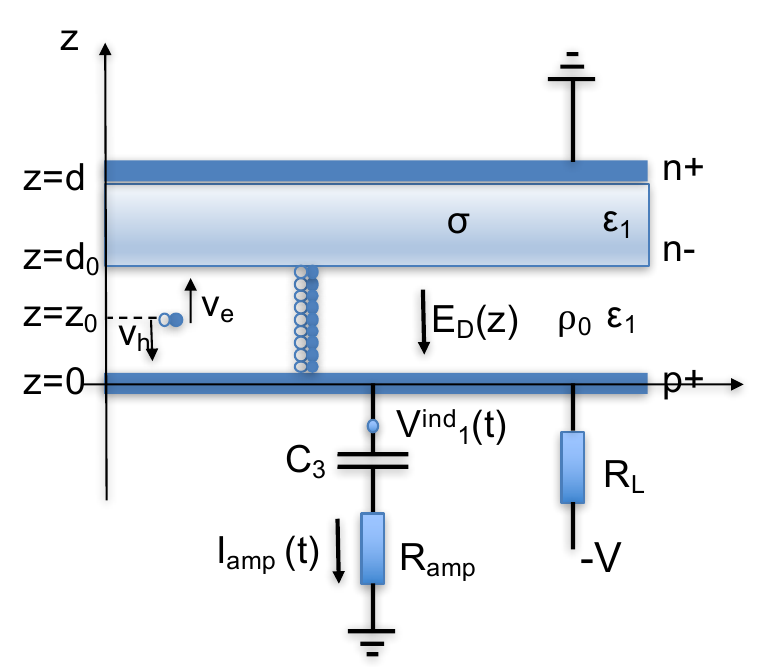, height=7cm}
\caption{A partially depleted silicon sensor assuming the movement of a single e-h pair and the movement of a continuous charge deposit, read out through capacitive coupling with an amplifier of input resistance $R_{amp}$. }
\label{example1}
\end{center}
\end{figure}

\noindent
To illustrate the above theorems we discuss a silicon sensor described in Example 5.3 from \cite{lutz} at an applied voltage $V$ that is smaller than the depletion voltage $V_{dep}$ (Fig.\,\ref{example1}). An n-type bulk material with doping concentration $N_D$ is equipped with a layer of highly doped p-type material. This p$^+$ layer is connected to the bias voltage $-V$ via a loading resistor $R_L$ and the signal is read out from this layer by an amplifier with input resistance $R_{amp}$ through a decoupling capacitor $C_3$. The other face of the sensor consists of a highly doped n$^+$ layer that is connected to ground. 
The depletion voltage $V_{dep}$ for the sensor and the thickness $d_0$ of the depleted layer for an applied voltage of $-V$  are  given by 
\beq
     V_{dep} = \frac{qN_Dd^2}{2\vep_1} \qquad   d_0 = d\,\sqrt{\frac{V}{V_{dep}}} \quad \mbox{for} \quad 0{<}V{<}V_{dep}
\eeq
where $q$ is the elementary charge and $\vep_1=\vep_r\vep_0$ is the dielectric permittivity of silicon. The static space charge density $\rho_0$ of the depleted layer and the conductivity $\sigma$ (the inverse of the volume resistivity)  of the un-depleted bulk layer are given by 
\beq
   \rho_0 = qN_D = \frac{2V_{dep}\vep_1}{d^2} \qquad   \sigma= q\mu_e N_D
\eeq
where $\mu_e$ is the electron mobility.

\subsection{Movement of the charges}

\noindent
For the following calculation we assume the n$^+$ and p$^+$ layers to have infinite conductivity and assume the boundary between the depleted and un-depleted layer at $z = d_0$ to be 'abrupt'. The drift field is defined by the potential $-V$ at $z=0$, a constant space-charge $\rho_0$ in $0<z<d_0$ as well as zero potential at $z=d_0$, which gives 
\beq
      E_D(z)  = -\frac{2V}{d_0} \left(1-\frac{z}{d_0} \right)
      \qquad 0{<}z{<}d_0
\eeq
The magnitude of the electric field decreases linearly from the value $E_D=-2V/d_0$ at $z=0$ to zero on the interface to the un-depleted layer at $z=d_0$. 
For low fields, the velocity of the electrons and holes is proportional to the electric field according to $v=\mu E$, so the movement of a single electron and a single hole deposited at $z=z_0$ is defined by the following differential equations
\beq
     \frac{d z_e(t)}{d t}   =   -\mu_e E_D(z_e(t)) \qquad    \frac{d z_h(t)}{d t}   =   \mu_h E_D(z_h(t)) \qquad z_e(0)=z_h(0)=z_0
\eeq
with the solution
\bea
      z_e(t) & = &  d_0-(d_0-z_0) e^{- t/\tau_e} \quad  \tau_e = \frac{d^2}{2\mu_eV_{dep}} \quad 0<t<\infty \\
            z_h(t) & = &  d_0-(d_0-z_0) e^{t/\tau_h} \quad  \tau_h = \frac{d^2}{2\mu_hV_{dep}} \quad 0<t<t_h
\eea
The holes take the time $t_h(z_0) = -\tau_h \ln \left( 1-\frac{z_0}{d_0} \right)$ to arrive at $z=0$, while the electrons take an infinite amount of time to arrive at $z=d_0$ since the electric field is zero at this position. The related velocities are:
\bea
     v_e(t) =  \frac{d z_e(t)}{d t} & = &  \frac{d_0-z_0}{\tau_e} e^{-t/ \tau_e}  \label{ve} \\
     v_h(t) =     \frac{ d z_h(t)}{d t} & = &  -\frac{d_0-z_0}{\tau_h} e^{t/\tau_h}  \, \Theta (t_h-t) \label{vh}
\eea

\subsection{Induced currents on the grounded electrode (Theorem 3)}

\begin{figure}[h]
\begin{center}
\epsfig{file=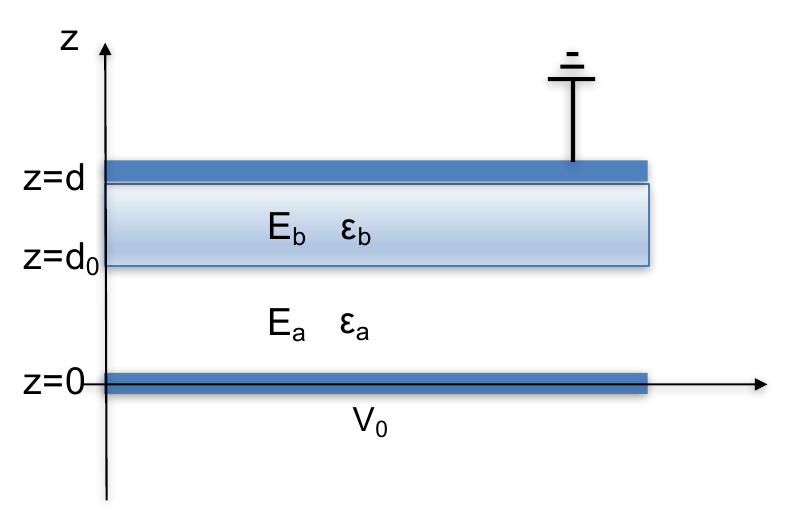, height=5cm}
\caption{Static electric fields in a parallel plate geometry with two layers of constant permittivity.}
\label{weightfield_static}
\end{center}
\end{figure}

\begin{figure}[h]
\begin{center}
\epsfig{file=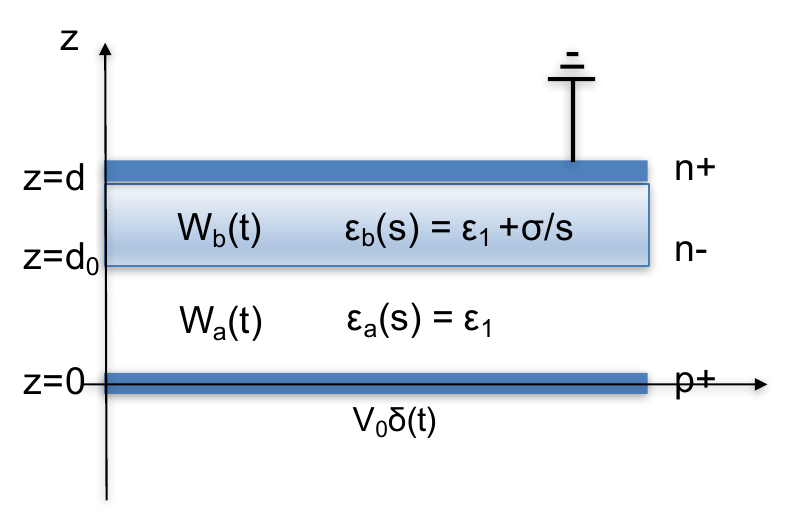, height=5cm}
\caption{Definition of the weighting field for calculation of the current induced on the grounded electrode.}
\label{example2}
\end{center}
\end{figure}

\begin{figure}[h]
\begin{center}
a)
\epsfig{file=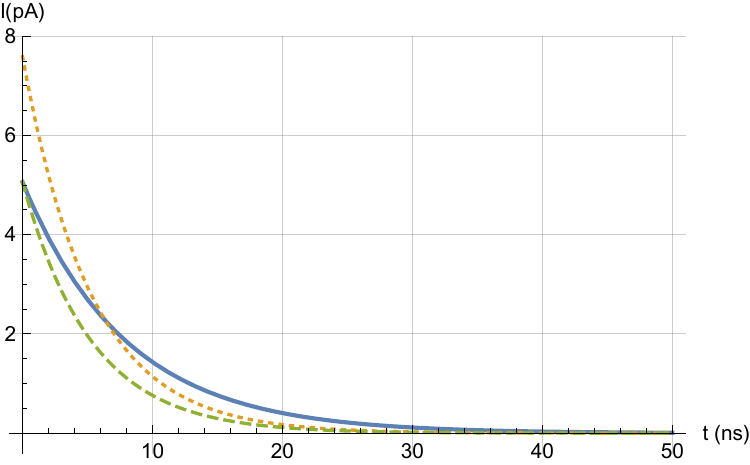, width=7.5cm}
b)
\epsfig{file=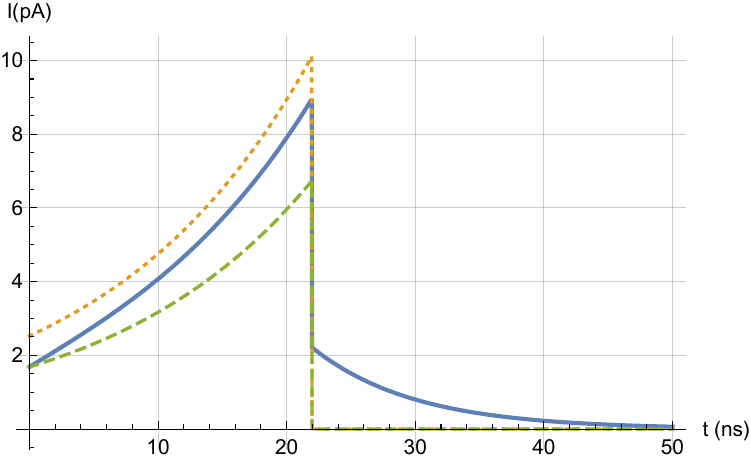, width=7.5cm}
c)
\epsfig{file=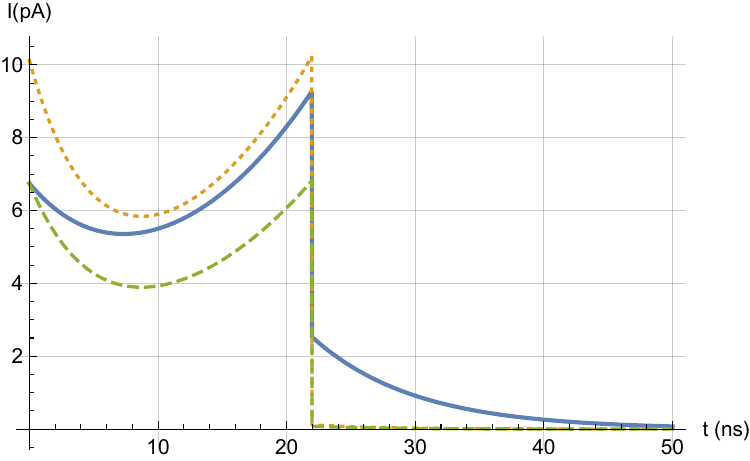, width=7.5cm}
\caption{Induced currents from a) a single electron b) a single hole c) an single electron+hole pair, for a sensor of 300\,$\mu$m thickness with a depletion voltage of 56.8\,V. The applied voltage is $V=-25.2$\,V resulting in a depleted region of $d_0=200\,\mu$m thickness. The e-h pair is deposited at $z_0=150\,\mu$m at $t=0$. The dotted line assumes zero  volume resistivity of the un-depleted layer and the dashed line assumes infinite volume resistivity.}
\label{single_charges}
\end{center}
\end{figure}

\begin{figure}[h]
\begin{center}
a)
\epsfig{file=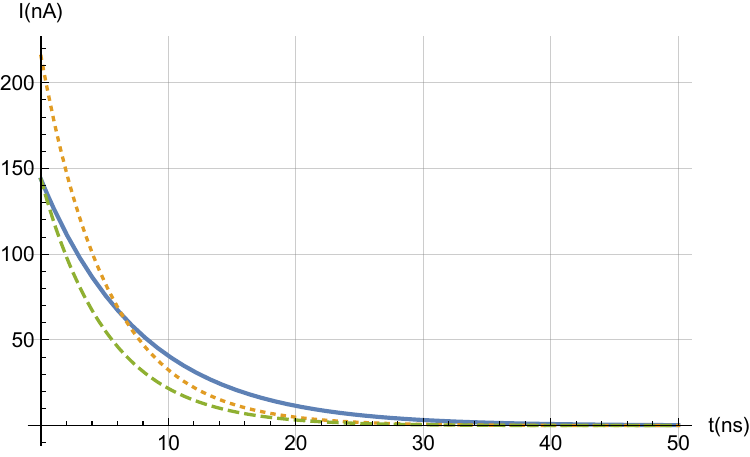, width=7.5cm}
b)
\epsfig{file=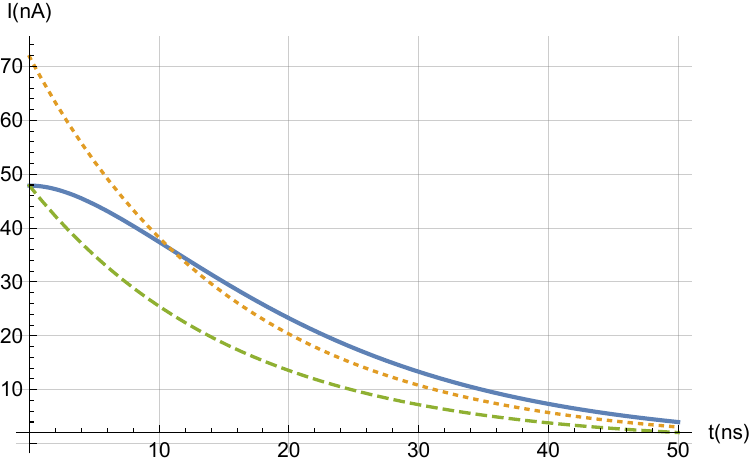, width=7.5cm}
c)
\epsfig{file=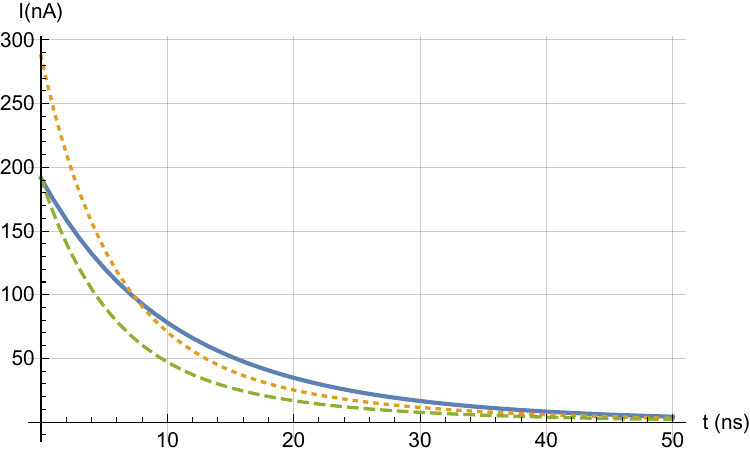, width=7.5cm}
\caption{Induced currents from a) a uniform distribution of electrons b) a uniform distribution of holes c) a uniform distribution of electrons and holes, for a sensor of 300\,$\mu$m thickness with a depletion voltage of 56.8\,V.  The applied voltage is $V=-25.2$\,V resulting in a depleted region of $d_0=200\,\mu$m thickness. The dotted line assumes zero  volume resistivity of the un-depleted layer and the dashed line assumes infinite volume resistivity.}
\label{linecharge}
\end{center}
\end{figure}

\noindent
To calculate the induced current for the case where the bottom electrode is grounded i.e. connected to ground through a very low impedance circuit, we have to apply a voltage pulse $V_0\delta(t)$ to the biased circuit in a TCAD simulation program and find the difference between this dynamic field and the static field as outlined in Theorem 3. For this analytic calculation we can derive the dynamic weighting field directly from a static solution together with the quasi-static approximation of Maxwell's equations. Fig. \ref{weightfield_static} shows a geometry with two layers of different permittivity and a static potential $V_0$ applied to the electrode at $z=0$. 
With the conditions $\vep_a E_a = \vep_b E_b$ and $E_ad_0+E_b(d-d_0)=V_0$ the fields can be calculated as 
\beq
     E_a = \frac{\vep_b V_0}{\vep_a(d-d_0) + \vep_b d_0} \qquad    E_b = \frac{\vep_a V_0}{\vep_a(d-d_0) + \vep_b d_0}
\eeq
The time dependent weighting vector for an application of a voltage pulse $V_0 \delta(t)$ as shown in Fig. \ref{example2} is then given by replacing $\vep_a$ with $\vep_1$, replacing $\vep_b$ by $\vep_1+\sigma/s$ and keeping $V_0(s)$ constant in these expressions and performing the inverse Laplace transform \cite{werner1}\cite{werner2}, which results in the desired weighting vector
\beq
   W_a(t)= \frac{V_0}{d} 
   \left(
   \delta(t) + \frac{d-d_0}{d_0}\,\frac{1}{\tau}e^{-t/\tau}
   \right) 
   \quad
    W_b(t)= \frac{V_0}{d} 
   \left(
   \delta(t) -\frac{1}{\tau}e^{-t/\tau}
   \right) 
   \qquad
      \tau= \frac{\vep_1d}{d_0 \sigma}  
\eeq
The time dependence signifies the reaction of the conductive medium. The electric fields from the primary charges will cause currents to flow in the conductive material, and this charge flow together with the primary charges are 'seen' by the metal electrodes, where signals are induced from both components. The delta function refers to the signals from the primary charge movement and the exponential part refers to the signals due to charge flow in the medium. In the limit of  infinite conductivity for the un-depleted layer we have $\tau=0$ and therefore
\beq \label{weighting_tau0}
   \lim_{\tau \rightarrow 0} W_a(t) = \frac{V_0}{d_0}\delta(t) \qquad   \lim_{\tau \rightarrow 0} W_b(t) = 0
\eeq
which is the correct static weighting field for the case of a parallel plate geometry with distance $d_0$. In case the layer has zero conductivity we have $\tau \rightarrow \infty$ and therefore
\beq \label{weighting_tauinf}
       \lim_{\tau \rightarrow \infty} W_a =    \lim_{\tau \rightarrow \infty} W_b = \frac{V_0}{d}  \delta(t)
\eeq
which is the correct static weighting for a parallel plate geometry with distance $d$.
The current induced by the single electron and the single hole starting from $z=z_0$ is then
\bea 
  i_e(t) & = &  \frac{q}{V_0} \int_0^t W_a(t-t')v_e(t')dt'  \no\\
  i_h(t)  & = &  -\frac{q}{V_0} \int_0^t W_a(t-t')v_h(t')dt'   \quad 0<t<t_h \label{convolution}\\
     & = &  -\frac{q}{V_0} \int_0^{t_h} W_a(t-t')v_h(t')dt'    \quad t > t_h \no
\eea
which evaluates to
\bea
    i_e(t, z_0)  & = &  q \frac{d_0-z_0}{d} \left[
            \frac{d-d_0}{d_0 (\tau-\tau_e)}\,\left(
            e^{-t/\tau} - e^{-t/\tau_e}
            \right)
            + \frac{1}{\tau_e} e^{-t/\tau_e}
     \right] \label{eqie} \\
        i_h(t,z_0) &= & q \frac{d_0-z_0}{d} \left[
      \frac{d-d_0}{d_0(\tau+\tau_h)}\left(
      e^{t/\tau_h} - e^{-t/\tau}
      \right)
      +\frac{1}{\tau_h}e^{t/\tau_h}
   \right] \qquad t<t_h  \label{eqih1} \\
   &= & q\frac{(d-d_0)(d_0-z_0)}{d_0d(\tau+\tau_h)}(e^{t_h(z_0)/\tau+t_h(z_0)/\tau_h}-1)\,e^{-t/\tau} \qquad t>t_h \label{eqih2} 
\eea
The charge induced by the electrons and the holes is given by
\beq
    Q_e=\int_0^\infty i_e(t) dt = q \left( 1-\frac{z_0}{d_0} \right) \qquad
     Q_h=\int_0^\infty i_h(t) dt = q \frac{z_0}{d_0} 
\eeq
so the total induced charge is $Q_{tot}= Q_e+Q_h=q$, as expected. 
In the limit of very small values of $\tau$, where the un-depleted layer acts like a perfect conductor, or very large values of $\tau$, where the un-depleted layer acts like a perfect insulator, the expressions for the induced currents are 
\beq
    \lim_{\tau \rightarrow 0} i_e(t) = q \frac{(d_0-z_0)}{d_0 \tau_e} \, e^{-t/\tau_e} =\frac{q}{d_0} v_e(t) \qquad
     \lim_{\tau \rightarrow 0} i_h(t) = q \frac{(d_0-z_0)}{d_0 \tau_h} \, e^{t/\tau_h} \Theta(t_h-t) =\frac{q}{d_0} v_h(t)
\eeq
\beq
        \lim_{\tau \rightarrow \infty} i_e(t) = q \frac{(d_0-z_0)}{d \tau_e} \, e^{-t/\tau_e} =\frac{q}{d} v_e(t) \qquad
            \lim_{\tau \rightarrow \infty} i_h(t) = q \frac{(d_0-z_0)}{d \tau_h} \, e^{t/\tau_h}\Theta(t_h-t)=\frac{q}{d_0} v_h(t)
\eeq
so the expressions with the usual static weighting fields are recovered. As an example we discuss a sensor with $d=300\,\mu$m, $N_D=8.3\times10^{11}$\,cm$^{-3}$ at a voltage of 25.2\,V and a single e-h pair placed at $z_0=150\,\mu$m. With the silicon parameters $\vep_1=11.8\vep_0$, $\mu_h=500$\,cm$^2$/Vs, $\mu_e=1500\,$cm$^2$/Vs we have $1/\sigma=5\,$k$\Omega$cm,  $V_{dep}=56.8$\,V, $d_0=200\,\mu$m. The time constants evaluate to $\tau=7.9$\,ns, $\tau_e=5.3$\,ns, $\tau_h=15.8$\,ns and the signals are shown in Fig.\,\ref{single_charges}. The electron is moving along the decreasing electric field towards the boundary at $z=d_0$ where the field is zero, so the decreasing velocity of the electron is responsible for the tail of the electron signal. The hole moves along the increasing electric field and the signal has an edge when the hole arrives at $z=0$ and stops moving. The tail of the signal is due to the charge movement in the un-depleted layer and the shape and magnitude depends on the volume resistivity.
\\ \\
Next we consider a continuous charge deposit approximating the ionization of a charged particle crossing the silicon sensor. We assume a linear charge density of $\lambda$\,[C/cm], so we have to replace $q$ by $dq = \lambda dz_0$ and integrate over $0<z_0<d_0$
\beq \label{eqIe}
   I_e(t)=\int_0^{d_0} \frac{\lambda}{q}i_e(t,z_0)dz_0 =\frac{d_0 \lambda}{2d(\tau-\tau_e)} 
   \left[
     (d-d_0)e^{-t/\tau}+(d_0\tau-d \tau_e)\frac{1}{
     \tau_e}e^{-t/\tau_e})
   \right]
   \eeq
\beq \label{eqIh}
   I_h(t)=\int_0^{d_0} \frac{\lambda}{q}i_h(t,z_0)dz_0 =\frac{d_0 \lambda}{2d(\tau-\tau_h)} 
   \left[
     (d-d_0)e^{-t/\tau}+(d_0\tau-d \tau_h)\frac{1}{
     \tau_h}e^{-t/\tau_h})
   \right]
   \eeq
The total induced charge is then
\beq
     Q^{ind}_{tot} =  \int_0^\infty [I_e(t)+I_h(t)] dt =  \frac{\lambda d_0}{2}+\frac{\lambda d_0}{2} = \lambda d_0
\eeq
as required. Here we have assumed that the charge deposited by the primary particle in the un-depleted layer has not diffused into the depletion region but has fully recombined. In the limit of very small values of $\tau$ or very large values of $\tau$ the expressions for the induced currents become
\beq
    \lim_{\tau \rightarrow 0} I_e(t) =  \frac{d_0 \lambda}{2\tau_e} \, e^{-t/\tau_e} \qquad
     \lim_{\tau \rightarrow 0} I_h(t) = \frac{d_0 \lambda}{ 2\tau_h} \, e^{-t/\tau_h} 
\eeq
\beq
        \lim_{\tau \rightarrow \infty} I_e(t) =  \frac{d_0^2 \lambda}{2d\tau_e} \, e^{-t/\tau_e} \qquad
            \lim_{\tau \rightarrow \infty} I_h(t) = \frac{d_0^2 \lambda}{ 2d\tau_h} \, e^{-t/\tau_h} 
\eeq
As an example we use a most probable number of 14200 e-h pairs in 200\,$\mu$m of silicon i.e. $\lambda = 14200q/200\mu$m, and the results are shown in Fig.\,\ref{linecharge}.
\\ \\
The entire calculation from above can actually also be performed in the Laplace domain. The convolutions in Eq.\,\ref{convolution} are simple products when working with the Laplace transforms of $W_a(t), v_e(t), v_h(t)$, so we have
\beq
     E_a(s) = V_0 \frac{(d+sd_0\tau)}{d_0d(1+s\tau)}
     \quad
     v_e(s) = \frac{d_0-z_0}{1+s\tau_e} 
     \quad
     v_h(s) = \frac{d_0-z_0}{1-s\tau_h} \,(1-e^{(1/\tau_h-s)t_h})
\eeq
\bea
   i_e(s, z_0) & = &  \phantom{-} \frac{q}{V_0} W_a(s)v_e(s) = 
   \phantom{-}q\frac{(d+sd_0\tau)}{d_0d(1+s\tau)} \frac{d_0-z_0}{1+s\tau_e}  \label{ie} \\
      i_h(s, z_0) & = &  -\frac{q}{V_0} W_a(s)v_h(s) =
     -q \frac{(d+sd_0\tau)}{d_0d(1+s\tau)} 
     \frac{d_0-z_0}{1-s\tau_h} \,(1-e^{(1/\tau_h-s)t_h(z_0)}) \label{ih}
\eea
Integrating over $z_0$ for the continuous charge deposit we have
\bea
    I_e(s) & = &  \int_0^{d_0} \frac{\lambda}{q} i_e(s, z_0) dz_0 = 
    \frac{d_0\lambda (d+sd_0\tau)}{2d(1+s\tau)(1+s\tau_e)} \\
    I_h(s) &  = &  \int_0^{d_0} \frac{\lambda}{q} i_h(s, z_0) dz_0   = \frac{d_0\lambda (d+sd_0\tau)}{2d(1+s\tau)(1+s\tau_h)} 
\eea
The expression $i_e(s, z_0), i_h(s, z_0), I_e(s), I_h(s)$ are the correct Laplace transforms of Eqs. \ref{eqie}, \ref{eqih1}, \ref{eqih2}, \ref{eqIe},  \ref{eqIh}.

\subsection{Induced voltage on the electrode connected by impedance elements (Theorem 1)}

\noindent
\begin{figure}[h]
\begin{center}
a)
     \epsfig{file=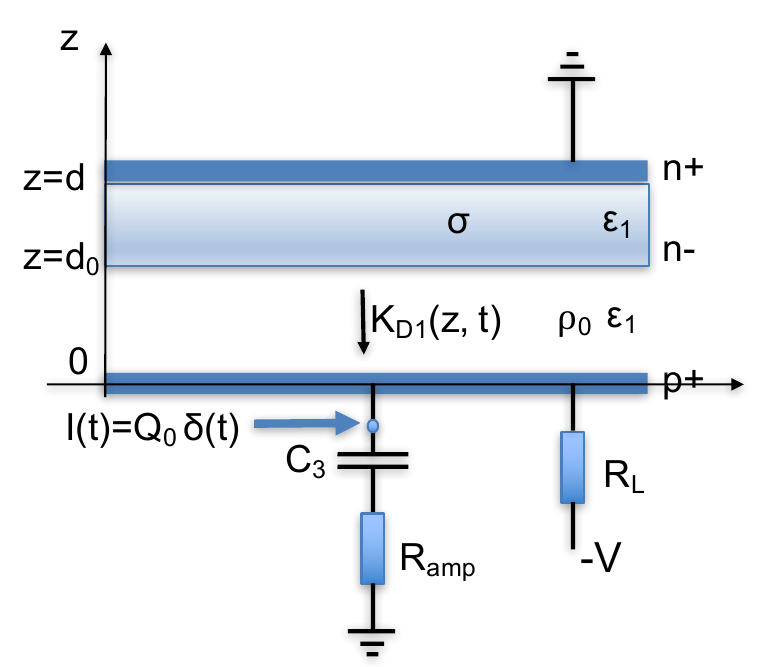, height=5cm}
b)
     \epsfig{file=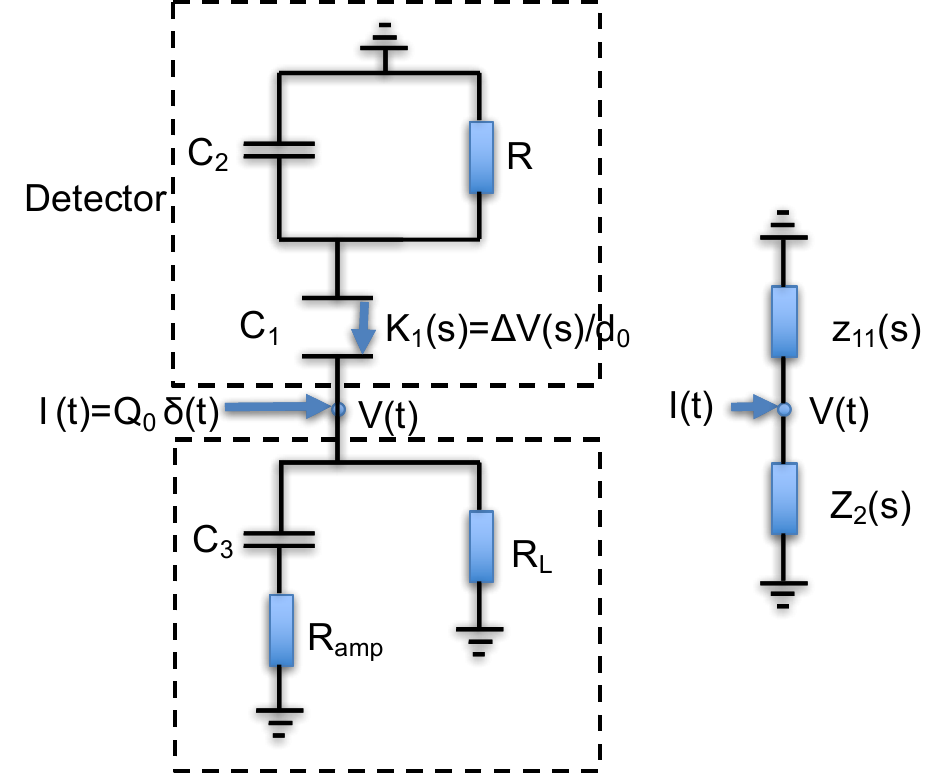, height=5cm}
\caption{a) Definition of the field $K_{D1}(z,t)$ b) Equivalent circuit for calculation of the weighting field $K_1(t)$.}
\label{equivalent_circuit}
\end{center}
\end{figure}
\begin{figure}[h] 
\begin{center}
     \epsfig{file=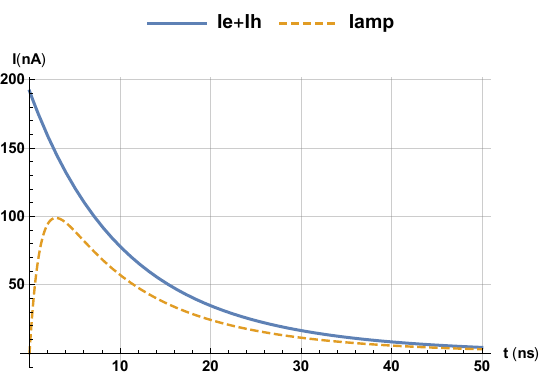, width=5cm}
\caption{The solid line shows the induced current from Fig. \ref{linecharge}c for the grounded electrode, while the dashed line shows the current in case the electrode is connected to an amplifier of finite input impedance (Eq. \ref{amp_current2}). The parameters are    $R= 50\,\Omega, C_1= 52\,$pF, $C_2=105\,$pF, $R_{amp}=50\,\Omega$ and $C_3=100$\,pF  for the geometry of Fig.\,\ref{example1}. The applied voltage is $V=-25.2$\,V resulting in a depleted region of $d_0=200\,\mu$m thickness.}
\label{amp_current}
\end{center}
\end{figure}

To illustrate Theorem 1 we calculate the induced voltage on the electrode when connected to the readout circuit and biasing network. To find the weighting field $K_1(z, t)$ we have to place a charge $Q(t)=Q_0 \Theta(t)$ or equivalently a current $I(t)=Q_0 \delta(t)$ on the electrode, as indicated in Fig.\,\ref{equivalent_circuit}a. 
As shown in \cite{werner2} the medium can be represented by an equivalent circuit. According to Eqs. 5 and 6 of \cite{werner2} the impedance 'matrix' $z_{11}$ of the electrode is given by
\beq
     1/z_{11} (s) =\frac{s}{V_0}\vep_1W_a(s)A =  \frac{\vep_1 s(1+\vep_1  s/\sigma)}{d_0+d\vep_1   s/\sigma} \,A
\eeq
where $A$ is the surface area of the detector. The equivalent circuit of the detector is therefore equal to a capacitor $C_1$ representing the depleted layer and a capacitor $C_2$ in parallel with a resistor $R$ representing the un-depleted layer (Fig.\,\ref{equivalent_circuit}b):
\beq
     C_1=\vep_1\frac{A}{d_0} \quad C_2=\vep_1\frac{A}{d-d_0} \quad R=\frac{1}{\sigma} \frac{d-d_0}{A}
     \qquad
     z_{11} (s) = \frac{1}{sC_1}+ \frac{R/(sC_2)}{R+1/(sC_2)}
\eeq
Connecting this equivalent circuit to the discrete readout elements, as shown in Fig.\,\ref{equivalent_circuit}, we can calculate the weighting field $K_1(t)$ by placing a current pulse $I(t)= Q_0\delta(t)$ i.e. $I(s)=Q_0$ on the electrode. The impedance $Z_2$ of the network connected to the electrode is  (Fig.\,\ref{equivalent_circuit}b)
\beq
     Z_2(s) = \frac{R_L(1+sC_3R_{amp})}{1+sC_3(R_L+R_{amp})}
\eeq
The potential $V(s)$ of the electrode and the electric field $K_1(s)=\Delta V/d_0$ inside the capacitor i.e in the depleted region, are then
\beq
        V(s) = Q_0 \frac{z_{11}(s)Z_2(s)}{z_{11}(s)+Z_2(s)} \qquad K_1(s) = \frac{Q_0}{d_0} \frac{1}{sC_1} \frac{Z_2(s)}{z_{11}(s)+Z_2(s)}
\eeq
This is the required weighting field $K_1(s)$, and the voltages induced by the electrons and holes for the setup in Fig.\,\ref{example1} are finally given by 
\beq
    u_e(s, z_0) = \frac{q}{Q_0}K_1(s) v_e(s, z_0) \qquad  u_h(s, z_0)=-\frac{q}{Q_0}K_1(s) v_h(s, z_0)
\eeq
From \cite{werner2} we know that this voltage can also be calculated by placing the currents $i_e(s)$ and $i_h(s)$ on the equivalent circuit, giving
\beq
       u_e(s, z_0) = \frac{Z_2(s)}{z_{11}(s)+Z_2(s)} z_{11}(s) i_e(s, z_0) \qquad   u_h(s, z_0) = \frac{Z_2(s)}{z_{11}(s)+Z_2(s)} z_{11} (s)i_h(s, z_0)
\eeq
Inserting the expressions from Eq.\,\ref{ie} and \ref{ih} shows that this is equivalent to the above expressions. The induced voltage for the uniform charge deposit is then
\beq
      V(s) =  \frac{z_{11}(s)Z_2(s)}{z_{11}(s)+Z_2(s)} \left[ I_e(s)+I_h(s) \right]
\eeq
Assuming $R_L$ to be chosen large enough to be negligible the current into the amplifier is 
\beq \label{amp_current2}
      I_{amp}(s) \approx  \frac{z_{11}(s)}{z_{11}(s)+Z_2(s)} \left[ I_e(s)+I_h(s) \right]
\eeq
As an example we use  a sensor area of $A=$1\,cm$^2$, which gives $R= 50\,\Omega, C_1= 52\,$pF, $C_2=105\,$pF. The applied voltage is $V=-25.2$\,V resulting in a depleted region of $d_0=200\,\mu$m thickness. For an amplifier input resistance $R_{amp}=50\,\Omega$ and a coupling capacitor of $C_3=100$\,pF we get the result shown in Fig.\,\ref{amp_current}. 

\section{Summary}

A formulation of extensions of the Ramo-Shockley theorem was presented. It allows the calculation of signals in detectors that contain non-linear materials of arbitrary permittivity and finite conductivity (volume resistivity), static space-charge as well as readout-electrodes  connected by an arbitrary impedance network. \\
The induced voltages and currents can be calculated by convolution of the velocity vectors of the moving charges with time dependent weighting fields or weighting vectors. These are defined by removing these charges and adding an infinitesimal voltage or current pulse to the electrode in question, while leaving all other applied potentials and impedance elements in place. The difference of these time dependent fields and the static drift field define the weighting fields or weighting vectors. \\
These theorems are very well suited for calculation of signals with TCAD device simulation programs, specifically for silicon sensors with un-depleted or partially depleted regions.  One can connect the biasing networks and impedances of the readout circuits and find the time dependent weighting field with numeric simulation programs. The statistics of primary ionization and the drift and diffusion of the charges  can then be calculated with a separate dedicated program and the movements of these charges are then simply convoluted with these weighting fields. \\
An analytic example for a simple silicon sensor was given. This example assumes uniform charge deposit along the track, it assumes an abrupt transition between depleted and un-depleted region and neglects charge carries moving from the un-depleted into the depleted region by diffusion. It can therefore only be used for benchmarking of TCAD simulations but not for direct comparison with measurements.
\\ \\
{\bf Acknowledgements} \\
I would like to thank Jan Hasenbichler, Heinrich Schindler and Ann Wang for important discussions. 


\end{document}